\documentclass[prl,aps,twocolumn,showpacs,floatfix,superscriptaddress]{revtex4}
\usepackage{graphicx}
\usepackage{amsmath}
\usepackage{amssymb}

\begin{document}

\title{Equilibration of a one-dimensional Wigner crystal}

\author{K. A. Matveev} 

\affiliation{Materials Science Division, Argonne National Laboratory,
  Argonne, IL 60439, USA}

\author{A. V. Andreev} 

\affiliation{Department of Physics, University of Washington, Seattle, WA
  98195, USA}

\author{M. Pustilnik}

\affiliation{School of Physics, Georgia Institute of Technology, Atlanta,
  Georgia 30332, USA}

\date{April 2, 2010}

\begin{abstract}

  Equilibration of a one-dimensional system of interacting electrons
  requires processes that change the numbers of left- and right-moving
  particles.  At low temperatures such processes are strongly suppressed,
  resulting in slow relaxation towards equilibrium.  We study this
  phenomenon in the case of spinless electrons with strong long-range
  repulsion, when the electrons form a one-dimensional Wigner crystal.  We
  find the relaxation rate by accounting for the Umklapp scattering of
  phonons in the crystal.  For the integrable model of particles with
  inverse-square repulsion, the relaxation rate vanishes.

\end{abstract}


\maketitle

The low energy properties of systems of interacting fermions in one
dimension are commonly described in the framework of the 
Luttinger liquid theory \cite{haldane, giamarchi}.  This theory
successfully predicted a number of interesting properties of
one-dimensional electron systems, such the power-law renormalization of
the tunneling density of states and impurity potential \cite{kane,
  furusaki}.  On the other hand, much recent interest in interacting
one-dimensional Fermi systems was focused on the phenomena not captured by
the Luttinger liquid theory \cite{pustilnik, lunde, fiete, meyer}.  One
example involves equilibration of a moving one-dimensional electron
liquid, which was recently shown to affect the conductance of quantum
wires \cite{rech} and drag between two wires \cite{mishchenko}.

In the case of weakly interacting one-dimensional electrons the physical
mechanism of equilibration was discussed in Ref.~\cite{micklitz}.  At low
temperature $T$ excitations of the system are particle-hole pairs near the
two Fermi points, with the typical energy $T\ll E_F$ and momentum
$T/v_F\ll p_F$. (Here $E_F$, $v_F$, and $p_F$ are the Fermi energy,
velocity and momentum of the system.)  In this regime, backscattering of a
right-moving electron near the Fermi level requires transfer of momentum
$2p_F$ to a large number of particle-hole pairs.  The most efficient such
process consists of a sequence of scattering events, in which a hole
passes from the left to the right Fermi point through the bottom of the
band.  Such processes are suppressed as $e^{-E_F/T}$ and, consequently,
equilibration of the chemical potentials of the right- and left-moving
electrons is a very slow process.

Equilibration of one-dimensional fermions beyond the weak interaction
regime is a more challenging problem.  In this case the description in the
language of particles and holes is no longer applicable, and according to
the Luttinger liquid theory the elementary excitations of the system are
bosons.  On the other hand, the Luttinger liquid theory does not
adequately describe particles near the bottom of the band, and is
therefore incapable of describing the equilibration processes.  In this
paper we show that this difficulty can be overcome in the limit of strong
long-range interactions.

More specifically, we consider a system of identical spinless particles of
mass $m$ described by the Hamiltonian of a general form
\begin{equation}
  \label{eq:H}
  H=\sum_l \frac{p_l^2}{2m} + \frac12\sum_{l,l'} V(x_l-x_{l'}).
\end{equation}
Here $x_l$ and $p_l$ are the coordinate and momentum of the $l$-th
particle, and $V(x)$ is the interaction potential.  In the limit of very
strong repulsion, the particles form a periodic chain with interparticle
distance $a=1/n$ determined by their density $n$.  In the case of Coulomb
repulsion, $V(x)=e^2/|x|$, such an arrangement is commonly referred to as
the Wigner crystal.  

At strong but finite repulsion, the particles can deviate from their
respective lattice sites, $x_l=la+u_l$, but the relative change of
interparticle distance remains small, $|u_l-u_{l'}|\ll|l-l'|a$.  To
leading order in the deviations $u_l$ the Hamiltonian (\ref{eq:H}) takes
the form
\begin{equation}
  \label{eq:H_0}
  H_0=\sum_l \frac{p_l^2}{2m} 
      + \frac14\sum_{l,l'} V_{l-l'}^{(2)}\,(u_l-u_{l'})^2,
\end{equation}
where we used the following notation for the $r$-th derivative of
the interaction potential
\begin{equation}
  \label{eq:V^(n)}
  V_l^{(r)}=\left.\frac{d^r V(x)}{dx^r}\right|_{x=la}.
\end{equation}
Elementary excitations of the harmonic chain (\ref{eq:H_0}) are phonons
characterized by quasimomentum $q$ (i.e., $u_l\propto e^{iql}$).  Their
frequencies are easily found by solving the classical equations of motion,
\begin{equation}
  \label{eq:omega_q}
  \omega_q^2=\frac{2}{m}\sum_{l=1}^\infty V^{(2)}_l [1-\cos(ql)].
\end{equation}
Provided that the interaction potential $V(x)$ falls off faster than
$1/|x|$ at large distances, the low-energy excitations of the system are
bosons with acoustic spectrum $\omega_q=s|q|$, in agreement with the
Luttinger liquid theory \cite{wigner-vs-luttinger}.  

Importantly, in the limit of strong interactions the harmonic
approximation (\ref{eq:H_0}) provides the full spectrum of elementary
excitations (\ref{eq:omega_q}), not limited by the restriction $|q|\ll 1$
imposed in the Luttinger liquid theory.  Another advantage of the Wigner
crystal model is that the weak interaction of phonons is naturally
described the anharmonic terms in the expansion of the Hamiltonian
(\ref{eq:H}) in the powers of the displacements $u_l$.  Scattering of
phonons resulting from these interactions leads to relaxation of their
distribution function to equilibrium.

At temperatures $T$ much lower than the Debye energy $\hbar\omega_\pi$,
the quasimomenta of the thermally excited phonons are small, $q\ll1$.  For
such phonons the Umklapp scattering is impossible, and phonon-phonon
collisions conserve the total quasimomentum $Q$ of the system.  As a
result, the equilibrium distribution of the phonons
\begin{equation}
  \label{eq:phonon_wind}
  N_q=\frac{1}{e^{\hbar(\omega_q-uq)/T}-1}
\end{equation}
is characterized by two parameters: the temperature $T$ and the velocity
$u$ of the phonon gas with respect to the lattice (see, e.g.,
Ref.~\cite{lifshitz}).

It is important to note that while being a good approximation,
conservation of quasimomentum is not exact.  Indeed, even at low
temperatures there is a finite occupation $N_\pi\sim
e^{-\hbar\omega_\pi/T}$ of phonon states near the edge $q=\pi$ of the
Brillouin zone, leading to a small probability of Umklapp processes.  As a
result, the total quasimomentum of the phonons relaxes as $\dot Q
=-Q/\tau$ with the time constant $\tau\propto e^{\hbar\omega_\pi/T}$.  For
distribution (\ref{eq:phonon_wind}) one has $Q\propto u$, and thus the
velocity $u$ acquires a time dependence, $\dot u=-u/\tau$.  Calculation of
the relaxation time $\tau$ is our main goal.

\begin{figure}[t]
 \resizebox{.48\textwidth}{!}{\includegraphics{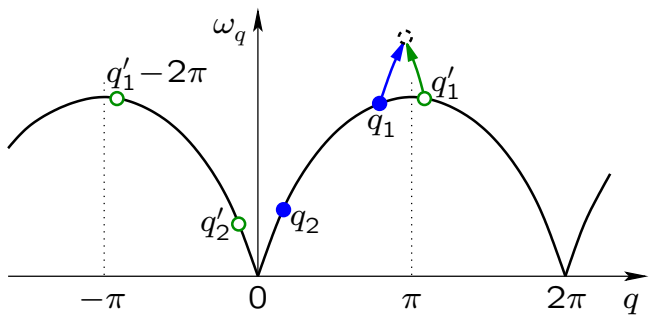}}
 \caption{\label{fig:spectrum} Umklapp scattering of two right-moving
   phonons (filled circles) into two left-moving phonons (open circles).
   The energies of the initial and final states are equal (see the dashed
   circle).  On the other hand, the sum of the quasimomenta defined to
   range in the first Brillouin zone $-\pi<q<\pi$, decreases by $2\pi$.}
\end{figure}

The microscopic mechanism of Umklapp scattering is illustrated in
Fig.~\ref{fig:spectrum}.  We assume that the phonon spectrum is concave,
which is the case for Coulomb repulsion.  Then the dominant process
involves scattering of a rare phonon with quasimomentum $q_1$ near the
boundary of the Brillouin zone by an acoustic phonon with energy
$\hbar\omega_{q_2}\sim T$.  As a result of such a collision the
quasimomentum of the high-energy phonon changes by $\delta q_1\sim T/\hbar
s$.  If the new quasimomentum $q_1'=q_1+\delta q_1$ is outside the
Brillouin zone ($-\pi$, $\pi$), the scattering involves Umklapp, and the
total quasimomentum changes by $\pm2\pi$.

In the course of  such scattering events the number of the rare phonons
near the edge of the Brillouin zone is conserved, while their momentum
changes by a small amount $\delta q\ll \pi$.  Thus these phonons essentially
diffuse in the momentum space, and the evolution of their distribution
function $P(q,t)$ can be described by the Fokker-Planck equation
\begin{equation}
  \label{eq:Fokker-Planck}
  \partial_t P=-\partial_q J, 
\quad
  J= A(q) P(q,t) - \frac12 \partial_q [B(q)P(q,t)].
\end{equation}
Here $J$ has the meaning of the probability current in momentum space, and
the functions $A(q)$ and $B(q)$ can be expressed in terms of the rate
$W_{q,q+\delta q}$ of the phonon transition from state $q$ to $q+\delta q$ as
\begin{equation}
   \label{eq:AB}
   A(q)=\sum_{\delta q} \delta q W_{q,q+\delta q},
\quad
   B(q)=\sum_{\delta q} (\delta q)^2W_{q,q+\delta q}.
\end{equation}
Below we will be using the Fokker-Planck equation (\ref{eq:Fokker-Planck})
to determine the behavior of the phonon distribution function $P$ in a
small vicinity of $q=\pi$.  This enables us to approximate $B(q)$ by
$B=B(\pi)$.  Furthermore, we will be interested in the case of weak
deviation from equilibrium, when the velocity of the phonon system is
small, $u\ll s$.  Thus we can approximate $A(q)$ by its value at
equilibrium, when the Boltzmann distribution $P(q)=e^{-\hbar\omega_q/T}$
has to solve the Fokker-Planck equation by nullifying the probability
current $J$.  Then, from (\ref{eq:Fokker-Planck}) we find
$A(q)=-B\hbar\omega'_q/2T$, where $\omega'_q=d\omega_q/dq$.

The phonon distribution function $P(q)$ is $2\pi$ periodic. In the first
Brillouin zone and away from its edges $q=\pm \pi$ it is given by
Eq.~(\ref{eq:phonon_wind}).  A periodic continuation of
(\ref{eq:phonon_wind}) results in a formal discontinuity at $q=\pi$.
Specifically, the phonon distribution function away from $q=\pi$ takes the
form
\begin{equation}
  \label{eq:boundary_conditions}
  P(q)=e^{-\hbar\omega_q/T} e^{\pm\pi\hbar u/T},
\quad
  \sqrt{\frac{T}{\hbar|\omega''_\pi|}}\ll \mp(q-\pi)\ll\pi.
\end{equation}
We now solve the Fokker-Planck equation (\ref{eq:Fokker-Planck}) with the
boundary conditions (\ref{eq:boundary_conditions}) to determine the
steady-state phonon distribution $P(q)$ at $|q-\pi|\lesssim
\sqrt{T/\hbar|\omega''_\pi|}$.  The solution is rather straightforward and
analogous to that for the distribution function of holes near the bottom
of the conduction band of weakly interacting electrons \cite{micklitz}.
It interpolates smoothly between the Boltzmann functions
(\ref{eq:boundary_conditions}) and corresponds to a finite but
exponentially small probability current
\begin{equation}
  \label{eq:J}
  J=uB\left(\frac{\pi\hbar^3|\omega''_\pi|}{2T^3}\right)^{1/2} 
    e^{-\hbar\omega_\pi/T}.
\end{equation}
The non-vanishing value of $J$ means that in unit time $(N/2\pi)J$ phonons
increase their momentum and leave the first Brillouin zone through point
$q=\pi$. (Here $N$ is the total number of electrons in the Wigner
crystal.) Each such event is an Umklapp process resulting in the decrease
of the quasimomentum $Q$ of the phonon system by $2\pi$.  We therefore
conclude that at non-zero phonon velocity $u$ the phonon scattering events
result in $\dot Q = - NJ$.  Comparing this result with the total
quasimomentum of the phonons $Q=\pi N u T^2/3s^3$, easily computed using
the distribution (\ref{eq:phonon_wind}), we find the relaxation rate
$\tau^{-1}=-\dot Q/Q$ in the form
\begin{equation}
  \label{eq:relaxation_rate}
  \tau^{-1}=3B\left(\frac{\hbar s}{T}\right)^3
                 \left(\frac{\hbar|\omega''_\pi|}{2\pi T}\right)^{1/2}
                 e^{-\hbar\omega_\pi/T}.
\end{equation}
As expected, the relaxation rate shows activated temperature dependence,
with activation temperature given by the Debye energy of the phonons
$\hbar\omega_\pi$.  However, to fully determine the temperature dependence
of the prefactor, one has to calculate the diffusion coefficient $B$ of
phonons in momentum space.

It is convenient to treat the scattering of phonons using the second
quantization of the Hamiltonian (\ref{eq:H}) whereby the displacements and
momenta of the particles are expressed in terms of the phonon destruction
and creation operators $b_q$ and $b_q^\dagger$ as
\begin{eqnarray}
  \label{eq:second_quantization}
  u_l &=& \sum_q \sqrt{\frac{\hbar}{2mN\omega_q}}\, (b_q+b_{-q}^\dagger)
  e^{iql},
\\
  p_l &=& -i\sum_q \sqrt{\frac{\hbar m\omega_q}{2N}}\, (b_q-b_{-q}^\dagger)
  e^{iql}.
\end{eqnarray}
The quadratic Hamiltonian (\ref{eq:H_0}) then takes the standard form
\begin{equation}
  \label{eq:H_0_second_quantized}
  H_0=\sum_q \hbar\omega_q(b_q^\dagger b_q+1/2).
\end{equation}
The coupling of phonons is described by the anharmonic corrections to
$H_0$, which are easily obtained by expanding the full Hamiltonian
(\ref{eq:H}) to higher orders in the displacements $u_l$.  To find the
leading contribution to phonon scattering, we will need to account only
for the cubic and quartic anharmonisms, $H\approx H_0+U^{(3)}+U^{(4)}$,
with the respective perturbations taking forms
\begin{widetext}
\begin{eqnarray}
  \label{eq:U^3}
  U^{(3)}&=&\frac{-i}{3\sqrt N}\left(\frac{\hbar}{2m}\right)^{3/2}
           \sum_{q_1,q_2}
           \frac{f_3(q_1,q_2)}{\sqrt{\omega_{q_1}\omega_{q_2}\omega_{q_1+q_2}}}\,
       (b_{q_1}+b_{-q_1}^\dagger)(b_{q_2}+b_{-q_2}^\dagger)(b_{-q_1-q_2}+b_{q_1+q_2}^\dagger),
\\
  \label{eq:U^4}
  U^{(4)}&=& \frac{\hbar^2}{48m^2N}
           \sum_{q_1,q_2,q_3}
           \frac{f_4(q_1,q_2,q_3)}
                {\sqrt{\omega_{q_1}\omega_{q_2}\omega_{q_3}\omega_{q_1+q_2+q_3}}}\,
                (b_{q_1}+b_{-q_1}^\dagger)(b_{q_2}+b_{-q_2}^\dagger)
                (b_{q_3}+b_{-q_3}^\dagger)(b_{-q_1-q_2-q_3}+b_{q_1+q_2+q_3}^\dagger).
\end{eqnarray}
\end{widetext}
Here the functions $f_3$ and $f_4$ are defined as
\begin{equation}
  \label{eq:f_3}
  f_3(q_1,q_2)= \sum_{l=1}^\infty V^{(3)}_l 
                 \{\sin[(q_1+q_2)l]-\sin(q_1l) -\sin(q_2l)\},
\end{equation}
and
\begin{eqnarray}
  \label{eq:f_4}
  f_4(q_1,q_2,q_3)&=& \sum_{l=1}^\infty V^{(4)}_l 
                     \{1-\cos(q_1l)-\cos(q_2l)
\nonumber\\
                  &&-\cos(q_3l)-\cos[(q_1+q_2+q_3)l]
\nonumber\\
                  &&+\cos[(q_1+q_2)l]+\cos[(q_1+q_3)l]
\nonumber\\
                  &&+\cos[(q_2+q_3)l]\}.
\end{eqnarray}

The rate of two-phonon scattering processes shown in
Fig.~\ref{fig:spectrum}, in which a phonon $q_1$ moves to the state
$q_1'$, is given by the golden rule expression
\begin{eqnarray}
  \label{eq:w}\hspace{-1em}
  W_{q_1,q_1'}&=&\frac{2\pi}{\hbar^2}
               \sum_{q_2,q_2'} |t_{q_1,q_2\to q_1',q_2'}|^2
               N_{q_2}(N_{q_2'}+1)
\nonumber\\
              &&\times
                 \delta_{q_1+q_2,q_1'+q_2'}
                 \delta(\omega_{q_1}+\omega_{q_2}
                        -\omega_{q_1'}-\omega_{q_2'}).
\end{eqnarray}
The scattering of two phonons $q_1$ and $q_2$ into $q_1'$ and $q_2'$, can
be accomplished in the first order in the quartic anharmonism $U^{(4)}$.
Alternatively, the same scattering process can be realized in second order
in the cubic anharmonism $U^{(3)}$.  Simple power counting shows that in
both cases the resulting amplitude is proportional to $\hbar^2$, i.e., one
has to account for both contributions.  The actual calculation is
straightforward and results in the scattering matrix element in the form
\begin{equation}
  \label{eq:matrix_element}
  t_{q_1,q_2\to q_1',q_2'}=\frac{\hbar^2}{m^3N}
                      \frac{\Lambda}
                      {(\omega_{q_1}\omega_{q_2}\omega_{q_1'}\omega_{q_2'})^{1/2}},
\end{equation}
where
\begin{eqnarray}
  \label{eq:Lambda}
  \Lambda&=&
                 -\frac{f_3(q_1,q_2)f_3(q_1',q_2')}
                       {\omega^2_{q_1+q_2}-(\omega_{q_1}+\omega_{q_2})^2}
+\frac{f_3(q_2,-q_1')f_3(q_1,-q_2')}
                       {\omega^2_{q_2-q_1'}-(\omega_{q_2}-\omega_{q_1'})^2}
\nonumber\\
               &&+\frac{f_3(q_1,-q_1')f_3(q_2,-q_2')}
                       {\omega^2_{q_2-q_2'}-(\omega_{q_2}-\omega_{q_2'})^2}
+\frac{m}{2}f_4(q_1,q_2,-q_1').
\end{eqnarray}

To find the diffusion constant $B(q_1)$ of phonons in momentum space, one
notes that the momentum $q_2$ is limited by the occupation number
$N_{q_2}$ in Eq.~(\ref{eq:w}), resulting in $|q_2|\sim|q_2'|\sim T/\hbar
s\ll \pi$.  Thus one can expand $\Lambda$ in powers of $\delta
q=q_1'-q_1=q_2-q_2'$.  The expansion starts with a quadratic term,
$\Lambda\propto (\delta q)^2$.  The proportionality constant depends on
the specific model of the interaction potential $V(x)$.  In the most
interesting case of Coulomb potential $V(x)=e^2/|x|$ our treatment is
complicated by the fact that the phonon speed $s$ diverges
logarithmically.  In practice, however, the Coulomb potential is always
screened at large distances by remote gates.  In this case, $\Lambda$ can
be found analytically.  For simplicity, we also limit ourselves to the
most important case of $q_1=\pi$ and find
\begin{equation}
  \label{eq:Lambda_Coulomb}
  \Lambda=\frac{21\zeta(3)}{32}\,\frac{me^2}{a^5}\,(\delta q)^2,
\end{equation}
where $\zeta(x)$ is the Riemann's zeta function.  Combining this result
with Eqs.~(\ref{eq:matrix_element}), (\ref{eq:w}), and (\ref{eq:AB}), we
find
\begin{equation}
  \label{eq:B}
  B=\chi\, T^5,
\quad
  \chi=\frac{21\pi^3\zeta(3)}{20}\, \frac{e^2}{\hbar^3 m^3 a^7 s^8}.
\end{equation}
One can now substitute this result into Eq.~(\ref{eq:relaxation_rate}) to
obtain the full expression for the relaxation rate of the phonon system in
a Wigner crystal.  In the specific case of Coulomb interaction, the Debye
frequency $\omega_\pi=(7\zeta(3)e^2/ma^3)^{1/2}$, parameter
$\omega_\pi''=-2\ln2(e^2/7\zeta(3)ma^3)^{1/2}$, and the speed of phonons
is $s=(2e^2\ln(d/a)/ma^3)^{1/2}$, where $d$ is the distance to the gate.

Although the expression (\ref{eq:B}) is derived for Coulomb interaction,
the temperature dependence $B\propto T^5$ is valid for any long-range
repulsive potential.  An interesting exception is the case of strong
inverse-square repulsion, $V(x)=\gamma/x^2$, with $\gamma\gg \hbar^2/m$.
This form of repulsive potential corresponds to the Calogero-Sutherland
model, which is exactly solvable due to the presence of an infinite number
of integrals of motion.  As a result, one expects that scattering of
excitations preserves their momenta, and no diffusion in momentum space
should be possible.  Indeed, we have been able to show that the
coefficient $\chi$ in Eq.~(\ref{eq:B}) vanishes for inverse-square
repulsion.  We have also verified that the expression (\ref{eq:Lambda})
for $\Lambda$, and thus the phonon scattering amplitude
(\ref{eq:matrix_element}), vanish in this case.  Scattering of a massive
particle off an acoustic phonon, Fig.~\ref{fig:spectrum}, was recently
discussed in the context of quantum decay of dark solitons in one
dimensional Bose systems \cite{gangardt}.  In analogy with our
observation, their decay rate vanishes in the integrable (Lieb-Liniger)
case.

The expression (\ref{eq:relaxation_rate}) for the relaxation rate of
phonon system in a one-dimensional Wigner crystal with the diffusion
constant (\ref{eq:B}) is the main result of this paper.  To illustrate its
significance we now briefly discuss the effect of the phonon relaxation
upon the conductance of a strongly-interacting quantum wire.  A more
rigorous treatment will be presented elsewhere.

The equilibration of phonons in a one-dimensional Wigner crystal is
analogous to electron equilibration in the limit of weak interactions.  In
the latter case, the excitations are holes/particles created by
transferring electrons to/from a Fermi point.  The momentum of each
excitation is thus measured from the nearest Fermi point.  Equilibration
processes include backscattering of holes near the bottom of the band.
Since the momenta of right- and left-moving holes are measured from
different Fermi points, each backscattering event changes the momentum of
the excitations by $2p_F=2\pi/a$, in analogy with the quasimomentum change
$\Delta Q=2\pi$ when a phonon in a Wigner crystal crosses the edge of the
Brillouin zone.  We thus conclude that $\dot Q$ can be interpreted as the
rate of backscattering of right-moving electrons, $\dot N^R=-\dot Q/2\pi$.

This relation provides for a way to observe the phonon equilibration by
measuring conductance of a strongly interacting wire connected to
non-interacting leads.  Negative value of $\dot N^R$ means that some of
the right-moving electrons entering the wire from the left return to the
same lead, thereby reducing the conductance \cite{lunde, rech, micklitz}.
Adapting the calculation of Ref.~\cite{micklitz} to the case strong
interactions, one finds
\begin{equation}
  \label{eq:conductance}
  G=\frac{e^2}{h}\left[1-L\frac{B}{a} 
                 \left(\frac{\hbar^3|\omega''_\pi|}{8\pi T^3}\right)^{1/2} 
                 e^{-\hbar\omega_\pi/T}\right].
\end{equation}
Although the correction to the conductance is exponentially small, it
grows with the length of the wire $L$.  The result (\ref{eq:conductance})
applies to relatively short wires, $L\ll l_{\rm eq}\propto
e^{\hbar\omega_\pi/T}$, c.f. Ref.~\cite{micklitz}. 

To summarise, we have studied the relaxation of the distribution function
of phonons in a one-dimensional spinless Wigner crystal.  Full
equilibration of phonons requires Umklapp processes, resulting in an
exponentially small relaxation rate.  The preexponential factor in the
resulting expression (\ref{eq:relaxation_rate}) scales as $T^{3/2}$.  In
the integrable case of inverse-square interactions the relaxation rate
vanishes.

The authors are grateful to D. M. Gangardt and A. Kamenev for helpful
discussions.  This work was supported by the U.S. Department of Energy
under Contract Nos. DE-AC02-06CH11357, DE-FG02-07ER46452, and
DE-FG02-06ER46311.

\end{document}